# The cosmic origin of quantum mechanics

Ding-Yu Chung

In this paper, quantum mechanics is explained by its cosmic origin. The base of quantum mechanics is the spontaneous tendency for a microscopic object to fractionalize instantly into quasistates and condense instantly quasistates. This quasistate is equivalent to the eigenfunction in the wavefunction. An object with the fractionalization-condensation is equivalent to the unitary wavefunction. Nonlocal operation is explicitly required to maintain communication among all quasistates regardless of distance during the fractionalization process. Interference effect is explicitly required for the condensation of quasistates. The collapse of the fractionalization-condensation is explicitly required when the fractionalization-condensation is disrupted externally by the interaction with environment and measurement or internally by transformation such as decay. The object without fractionalization-condensation is the classical object without nonlocality-interference. The cosmic origin of quantum mechanics is derived from the cyclic fractionalization-condensation in the cyclic universe, consisting of the unobservable cosmic vacuum and the observable universe. The cyclic fractionalization-condensation allows quasistates to appear cyclically rather than simultaneously. The unobservable cosmic vacuum involves the gradual cyclic fractionalization-condensation between the high energy eleven dimensional spacetime and low energy four dimensional spacetime. The observable universe involves the drastic cyclic fractionalization-condensation consisting of the cosmic instant fractionalization (the big bang) into various dimensional particles and the expansion-contraction by mostly cosmic radiation and gravity. The big bang (cosmic instant fractionalization) in the observable universe leads to the microscopic instant fractionalization-condensation (the standard quantum mechanics) that allows all quasistates from an object to appear simultaneously. The big bang generates missing space dimensions that are filled by internal space dimensions with internal symmetries. The products of the big bang are the four spacetime dimensional ordinary (baryonic) matters and fields with seven dimensional internal space (non-spacetime), the exotic dark matters with higher spacetime dimensions, and the four spacetime dimensional gravity, cosmic radiation, and ordinary vacuum without internal space. The cosmic connection between the cosmic vacuum and the ordinary vacuum becomes active, when the spacetime dimensions of the cosmic vacuum and the ordinary vacuum are compatible. The active cosmic connection allows the observable universe to have expansion input or contraction input from the expansion or contraction of the cosmic vacuum. The results are the inflation before the big bang, the late-time cosmic accelerating expansion, the cosmic accelerating contraction, and the final cosmic accelerating contraction. Through different routes, both the unobservable cosmic vacuum and the observable universe achieve the cyclic fractionalization-condensation.

## 1. Introduction

The study of semi-conductor crystals at low temperature shows the fractionalization of electron in the form of a separation of spin and charge or literal fractionalization of electric charge [1]. H. J. Maris also found the evidences for quasi-electrons in liquid helium [2]. In this paper, this spontaneous tendency for a microscopic object to fractionalize is the base to interpret quantum mechanics.

The three most important non-classical features in quantum mechanics are the nonlocal operation, the interference effect, and the collapse of wavefunction. In this paper, the base of quantum mechanics is the spontaneous tendency for a microscopic object to fractionalize instantly into quasistates and condense instantly quasistates. This quasistate is equivalent to the eigenfunction in the wavefunction. An object with the fractionalization-condensation is equivalent to the unitary wavefunction. Nonlocal operation is explicitly required to maintain communication among all quasistates regardless of distance during the fractionalization process. Interference effect is explicitly required for the condensation of quasistates. The collapse of the fractionalization-condensation is explicitly required when the fractionalization-condensation is disrupted. The elimination of the fractionalization-condensation leads to the object without nonlocality-interference.

The cosmic origin of quantum mechanics is derived from the cyclic fractionalization-condensation in the cyclic universe, consisting of the unobservable cosmic vacuum and the observable universe. The cyclic fractionalization-condensation allows quasistates to appear cyclically rather than simultaneously. The big bang (cosmic instant fractionalization) in the observable universe leads to the microscopic instant fractionalization-condensation that allows all quasistates from an object to appear simultaneously. The microscopic instant fractionalization-condensation is the standard quantum mechanics.

The cosmic connection between the unobservable cosmic vacuum and the observable ordinary vacuum becomes active, when the spacetime dimensions of the cosmic vacuum and the ordinary vacuum are compatible. The active cosmic connection allows the observable universe to have expansion input or contraction input from the expansion or contraction of the cosmic vacuum.

In the Section 2, the fractionalization-condensation will be discussed. In the Section 3, the unobservable cosmic vacuum will be discussed. The Section 4 covers the observable universe. The Section 5 reviews the ordinary universe. In the Section 6, the four periods of the active cosmic connection will be discussed.

## 2. The Fractionalization-condensation

An object has a spontaneous tendency to fractionalize instantly into quasistates and condense instantly quasistates. This quasistate is equivalent to the eigenfunction in the wavefunction.  An object with the fractionalization-condensation is equivalent to the wavefunction.   All quasitates during fractionalization can communicate with one another to preserve the integrity of the object regardless of distance, implying nonlocal operation.  The integrity of an object is equivalent to the unitary in the wavefunction.  In the wavefunction, all eigenfunctions can communicate with one another.  There is no time in the wavefunction at a fundamental level.   The combination of the fractionalization and the integrity leads to nonlocality in quantum mechanics.

At the same time, all quasitates during condensation overlap with one another because they are originally from one source.   The combination of the condensation and the integrity results in interference effect in quantum mechanics.

The disruption of the fractionalization-condensation is equivalent to the collapse of wavefunction.  When the fractionalization-condensation is disrupted externally by the interaction with environment  [3] or measurement, or is disrupted by internal transformation, such as decay, an object loses the fractionalization-condensation (nonlocality-interference).  Therefore, when an object is isolated and stable, the object has nonlocality-interference.   When an object is not isolated and stable, the object loses nonlocality-interference, and becomes a classical object instead of quantum object.

The cosmic origin of quantum mechanics is derived from the cyclic fractionalization-condensation in the cyclic universe, consisting of the unobservable cosmic vacuum and the observable universe.  The cyclic fractionalization-condensation allows quasitates to appear cyclically rather than simultaneously.  The unobservable cosmic vacuum involves the gradual cyclic fractionalization-condensation between the high energy eleven dimensional spacetime and low energy four dimensional spacetime.  The observable universe involves the drastic cyclic fractionalization-condensation consisting of the cosmic instant fractionalization (the big bang) into various dimensional particles and the expansion-contraction by mostly cosmic radiation and gravity.  During the big bang, the eleven dimensional supermembranes fractionalize into various dimensional particles instantly.  The big bang (cosmic instant fractionalization) leads to the microscopic instant fractionalization-condensation that allows all quasitates from an object to appear simultaneously.  The microscopic instant fractionalization-condensation is the standard quantum mechanics.

### 3.     *The Unobservable Cosmic Vacuum*

The unobservable cosmic vacuum starts with the superposition of dimensions in the eleven dimensional homogeneous Planck supermembranes, and undergoes the gradual fractionalization-condensation, resulting in the gigantic slow cosmic oscillation (expansion-contraction) between the high energy small eleven dimensional spacetime and the low energy large four dimensional spacetime.

A supermembrane can be described as two dimensional object that moves in an eleven dimensional space-time [4]. This supermembrane can be converted into the ten dimensional superstring with the extra dimension curled into a circle to become a closed superstring. The cosmic vacuum state is the wavefunction of this eleven-dimensional Planck supermembrane. The Planck wavefunction is the superposition of dimensions from eleven to four dimensional spacetime with decreasing energy and increasing size. In the Planck wavefunction, each space dimension can be described by a fermion and a boson as the following hierarchy:

$F_4\ B_4\ F_5\ B_5\ F_6\ B_6\ F_7\ B_7\ F_8\ B_8\ F_9\ B_9\ F_{10}\ B_{10}\ F_{11}\ B_{11}$

where B and F are boson and fermion in each spacetime dimension. The probability to transforming a fermion into its boson partner in the adjacent dimension is same as the fine structure constant, $\alpha$, the probability of a fermion emitting or absorbing a boson. The probability to transforming a boson into its fermion partner in the same dimension is also the fine structure constant, $\alpha$. This hierarchy can be expressed in term of the dimension number, D,

$$M_{D-1,\ B} = M_{D,\ F}\ \alpha_{D,\ F}, \quad (1)$$

$$M_{D,\ F} = M_{D,\ B}\ \alpha_{D,\ B}, \quad (2)$$

where $M_{D,\ B}$ and $M_{D,\ F}$ are the masses for a boson and a fermion, respectively, and $\alpha_{D,\ B}$ or $\alpha_{D,F}$ is the fine structure constant, which is the ratio between the energies of a boson and its fermionic partner. All fermions and bosons are related by the order $1/\alpha$.

The cosmic vacuum is a dynamic vacuum. Through the gradual cyclic fractionalization-condensation, the cosmic vacuum undergoes a gigantic slow cosmic oscillation between the high-energy small eleven dimensional spacetime and the low-energy large four dimensional spacetime.

The energy as the cosmological constant in the vacuum state is (mass)$^4$ where the mass is the mass for the symmetry breaking. The vacuum energy for the cosmic vacuum is, therefore, equal to $M_p^4$ where $M_p$ is the Planck mass. Assuming a logarithm relation, the corresponding Planck time in the cosmic vacuum is equal to $T_p^{1/4}$ where $T_p$ is the Planck time (5.39 x$10^{-44}$ second). The Planck time in the cosmic vacuum is, therefore, equal to 1.52 x$10^{-11}$ second. Since fractionalization rate is related to the Planck time, the fractionalization rate in the cosmic vacuum is one fractional particle per 1.52 x$10^{-11}$ second per parent Planck supermembrane. It takes 2.1 x $10^{-9}$ second (the inflation time) to fractionalize the eleven dimensional Planck supermembrane to the eleven dimensional fermion. It takes 27.6 billion years (the half cycle time of the universe) to fractionalize from the eleven dimensional Planck supermembrane to the four dimensional bosons.

## 4. *The Observable Universe*

The observable universe starts with the superposition of dimensions in the eleven dimensional heterogeneous Planck supermembranes with unequal sets (lepton-quark) of spacetime (Fig. 1). Unequal sets of spacetime induce entropy, and cause cosmic instant fractionalization into particles with different spacetime dimensions, resulting in the big bang. The big bang generates missing space dimensions that are filled by internal space dimensions with internal symmetries. The products of the big bang are the four spacetime dimensional ordinary (baryonic) matters and fields with seven dimensional internal space (non-spacetime), the exotic dark matters with higher spacetime dimensions, and the four spacetime dimensional gravity, cosmic radiation, and ordinary vacuum without internal space. The current density of the ordinary matter is 0.012 - 0.041, while the density of the exotic dark matter is 0.1 - 0.5 [5]. Since the ordinary matter is one of the eight possible dimension quasistates, the ratio of the ordinary matter to the exotic dark matter agrees with the observed value.

The mechanism for the cyclic fractionalization-condensation of the observable universe is driven by mostly cosmic radiation and gravity, corresponding to the expansion and the contraction in the cyclic fractionalization-condensation. Cosmic radiation maximizes the distance between any two objects as in the expansion, while gravity minimizes the distance between any two objects as in the contraction. The interactions of cosmic radiation and gravity toward all materials are indiscriminate. Everything can be involved in the same expansion and the contraction. Both cosmic radiation and gravity follow the property of spacetime. Ubiquitous cosmic radiation represents spacetime, and gravity is the curvature of four dimensional spacetime. Cosmic radiation and gravity do not need internal space.

## 4. *The Ordinary Universe*

The ordinary universe includes the four spacetime dimensional ordinary (baryonic) matters and fields with seven dimensional internal space (non-spacetime) and four spacetime dimensional cosmic radiation, gravity, and ordinary vacuum without internal space. The seven dimensional internal space is the base for the ordinary gauge bosons and the periodic table of ordinary elementary particles as described in details in Reference 6. It is briefly reviewed here.

The permanent lepton-quark composite state requires the violations of symmetries (CP and P) in spacetime and the internal symmetry acting on the internal space. CP nonconservation is required to distinguish the lepton-quark composite state from the CP symmetrical cosmic radiation that is absence of the lepton-quark composite state. P nonconservation is required to achieve chiral symmetry for massless leptons (neutrinos), so some of the dimensional fermions can become leptons (neutrinos). Various internal symmetry groups are require to organize leptons, quarks, black hole particles, and force fields.  The force fields include the long-range massless force to bind leptons and quarks, the short-range force to bind quarks, the short-range interactions for the flavor change among quarks and leptons, and the short-range interaction for the eleven dimensional particles inside black hole.

The ordinary universe starts with the superposition of dimensions in the eleven dimensional heterogeneous Planck supermembranes with unequal sets (lepton-quark) of spacetime (Fig. 1).  The lepton-quark composite state consists of two sets of seven dimensional internal dimensional space.  The internal space mostly for leptons becomes the dominating internal space, while the internal space mostly for quarks is hidden.   Finally, the ordinary gauge bosons, leptons, and quarks absorb the common scalar field of four dimensional spacetime to acquire the common label of four dimensional spacetime through Higgs mechanism.   The results are the ordinary gauge bosons (Table 1) and the periodic table of ordinary elementary particles (Table 2).

For the ordinary universe, the seven internal space dimensions are arranged in the same way as the spacetime dimensions in the cosmic vacuum.

$$F_5\ B_5\ F_6\ B_6\ F_7\ B_7\ F_8\ B_8\ F_9\ B_9\ F_{10}\ B_{10}\ F_{11}\ B_{11}$$

where B and F are boson and fermion in each spacetime dimension. The ordinary gauge bosons can be derived from Eqs. (1) and (2).   Assuming $\alpha_{D,B} = \alpha_{D,F}$, the relation between the bosons in the adjacent dimensions, then, can be expressed in term of the dimension number, D,

$$M_{D-1, B} = M_{D, B}\ \alpha^2_D, \qquad (3)$$

where D= 6 to 11, and $E_{5,B}$ and $E_{11,B}$ are the energies for the dimension five and the dimension eleven, respectively.

The lowest  energy is the Coulombic field, $E_{5,B}$

$$\begin{aligned} E_{5, B} &= \alpha\ M_{6,F} \\ &= \alpha\ M_e, \end{aligned} \qquad (4)$$

where $M_e$ is the rest energy of electron, and $\alpha = \alpha_e$, the fine structure constant for the magnetic field.  The bosons generated are called "dimensional bosons" or "$B_D$". Using only $\alpha_e$, the mass of electron, the mass of $Z^0$, and the number of extra dimensions (seven), the masses of $B_D$ as the ordinary gauge boson can be calculated as shown in Table 1.

**Table 1.** The Energies of the Dimensional Bosons
$B_D$ = dimensional boson, $\alpha = \alpha_e$

| $B_D$ | $M_D$ | GeV | Ordinary gauge boson | Interaction, internal symmetry |
|---|---|---|---|---|
| $B_5$ | $M_e \alpha$ | $3.7 \times 10^{-6}$ | A | electromagnetic, U(1) |
| $B_6$ | $M_e/\alpha$ | $7 \times 10^{-2}$ | $\pi_{1/2}$ | strong, SU(3) |
| $B_7$ | $M_6/\alpha_w^2 \cos\theta_w$ | 91.177 | $Z_L^0$ | weak (left), $SU(2)_L$ |
| $B_8$ | $M_7/\alpha^2$ | $1.7 \times 10^6$ | $X_R$ | CP (right) nonconservation, $U(1)_R$ |
| $B_9$ | $M_8/\alpha^2$ | $3.2 \times 10^{10}$ | $X_L$ | CP (left) nonconservation, $U(1)_L$ |
| $B_{10}$ | $M_9/\alpha^2$ | $6.0 \times 10^{14}$ | $Z_R^0$ | weak (right), $SU(2)_R$ |
| $B_{11}$ | $M_{10}/\alpha^2$ | $1.1 \times 10^{19}$ | $G_b$ | black hole, large N color field |

      In Table 1, $\alpha_w$ is not same as $\alpha$ of the rest, because there is symmetry group mixing between $B_5$ and $B_7$ as the symmetry mixing in the standard theory of the electroweak interaction, and $\sin\theta_w$ is not equal to 1. As shown in Reference 6, $B_5$, $B_6$, $B_7$, $B_8$, $B_9$, and $B_{10}$ are A (massless photon), $\pi_{1/2}$, $Z_L^0$, $X_R$, $X_L$, and $Z_R^0$, respectively, responsible for the electromagnetic field, the strong interaction, the weak (left handed) interaction, the CP (right handed) nonconservation, the CP (left handed) nonconservation, and the P (right handed) nonconservation, respectively. The calculated value for $\theta_w$ is $29.69^0$ in good agreement with $28.7^0$ for the observed value of $\theta_w$ [7].

      The calculated energy for $B_{11}$ ($G_b$, black hole gluon) is $1.1 \times 10^{19}$ GeV in good agreement with the Planck mass, $1.2 \times 10^{19}$ GeV. $G_b$ is the eleven dimensional gauge boson for the interaction inside the event horizon of the black hole. The confinement of all particles inside black hole is similar to the confinement of all quarks inside hardon, so the internal symmetry is based on large N color [8]. Black hole gluon corresponds to gluon, and black hole gluino (the eleven dimensional fermion), the supersymmetrical partner, corresponds to quark. Therefore, inside the horizon of black hole, the gravitational force goes zero, and the large N color field appears. The eleven dimensional black hole gluon and gluino are surrounded by the four dimensional gravity.

      These eleven dimensional black hole gluons and gluinos are still in internal space (color) whose internal symmetry generates large N color force field. They are not eleven dimensional Planck supermembranes that hold together by the tangle of D-branes and strings. The event horizon is the dividing line between the four dimensional spacetime labeled force and the internal space labeled force. As described later, all black hole universe is the end of the cosmic cycle, and all Planck supermembrane is the beginning of the cosmic cycle.

The calculated masses of all ordinary gauge bosons are in good agreement with the observed values. Most importantly, the calculation shows that exactly seven extra dimensions are needed for all fundamental interactions.

The model for leptons and quarks is shown in Fig. 1. The periodic table for ordinary elementary particles is shown in Table 2.

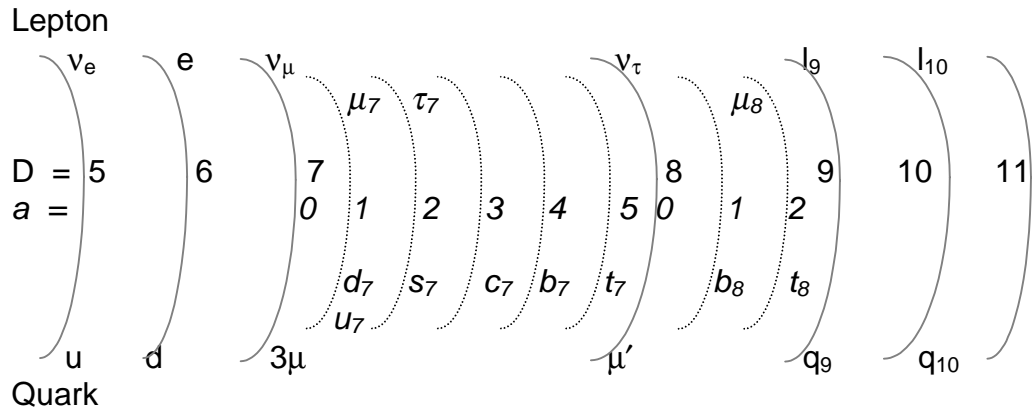

**Fig. 1.** Ordinary leptons and quarks in the dimensional orbits
D = dimensional number, a = auxiliary dimensional number

**Table 2.** The Periodic Table of ordinary elementary particles
D = dimensional number, a = auxiliary dimensional number

| D  | A = 0 | 1 | 2 | a = 0 | 1 | 2 | 3 | 4 | 5 | Boson |
|----|-------|---|---|-------|---|---|---|---|---|-------|
|    | Lepton |   |   | Quark |   |   |   |   |   | Boson |
| 5  | $l_5 = \nu_e$ |   |   | $q_5 = u = 3\nu_e$ |   |   |   |   |   | $B_5 = A$ |
| 6  | $l_6 = e$ |   |   | $q_6 = d = 3e$ |   |   |   |   |   | $B_6 = \pi_{1/2}$ |
| 7  | $l_7 = \nu_\mu$ | $\mu_7$ | $\tau_7$ | $q_7 = 3\mu$ | $u_7/d_7$ | $s_7$ | $c_7$ | $b_7$ | $t_7$ | $B_7 = Z_L^0$ |
| 8  | $l_8 = \nu_\tau$ | $\mu_8$ |   | $q_8 = \mu'$ | $b_8$ | $t_8$ |   |   |   | $B_8 = X_R$ |
| 9  | $l_9$ |   |   | $q_9$ |   |   |   |   |   | $B_9 = X_L$ |
| 10 | $F_{10}$ |   |   |   |   |   |   |   |   | $B_{10} = Z_R^0$ |
| 11 | $F_{11}$ |   |   |   |   |   |   |   |   | $B_{11} = G_b$ |

D is the dimensional orbital number for the seven extra space dimensions. The auxiliary dimensional orbital number, a, is for the seven extra auxiliary space dimensions, mostly for subquarks. All ordinary gauge bosons, leptons, and subquarks are located on the seven dimensional orbits and seven auxiliary orbits. Most leptons are dimensional fermions, while all quarks are the sums of subquarks.

The fermion mass formula for massive leptons and quarks is derived from Reference 6 as follows.

$$M_{F_{D,a}} = \sum M_{F_{D,0}} + M_{AF_{D,a}}$$

$$= \sum M_{F_{D,0}} + \frac{3}{2} M_{B_{D-1,0}} \sum_{a=0}^{a} a^4 \quad (5)$$

$$= \sum M_{F_{D,0}} + \frac{3}{2} M_{F_{D,0}} \alpha_D \sum_{a=0}^{a} a^4$$

Each fermion can be defined by dimensional numbers (D's) and auxiliary dimensional numbers (a's). The compositions and calculated masses of ordinary leptons and quarks are listed in Table 3.

**Table 3.** The Compositions and the Constituent Masses of Ordinary Leptons and Quarks
D = dimensional number and a = auxiliary dimensional number

|  | $D_a$ | Composition | Calc. Mass |
|---|---|---|---|
| Leptons | $D_a$ for leptons |  |  |
| $\nu_e$ | $5_0$ | $\nu_e$ | 0 |
| e | $6_0$ | e | 0.51 MeV (given) |
| $\nu_\mu$ | $7_0$ | $\nu_\mu$ | 0 |
| $\nu_\tau$ | $8_0$ | $\nu_\tau$ | 0 |
| $\mu$ | $6_0 + 7_0 + 7_1$ | $e + \nu_\mu + \mu_7$ | 105.6 MeV |
| $\tau$ | $6_0 + 7_0 + 7_2$ | $e + \nu_\mu + \tau_7$ | 1786 MeV |
| $\mu'$ | $6_0 + 7_0 + 7_2 + 8_0 + 8_1$ | $e + \nu_\mu + \mu_7 + \nu_\tau + \mu_8$ | 136.9 GeV |
| Quarks | $D_a$ for quarks |  |  |
| u | $5_0 + 7_0 + 7_1$ | $u_5 + q_7 + u_7$ | 330.8 MeV |
| d | $6_0 + 7_0 + 7_1$ | $d_6 + q_7 + d_7$ | 332.3 MeV |
| s | $6_0 + 7_0 + 7_2$ | $d_6 + q_7 + s_7$ | 558 MeV |
| c | $5_0 + 7_0 + 7_3$ | $u_5 + q_7 + c_7$ | 1701 MeV |
| b | $6_0 + 7_0 + 7_4$ | $d_6 + q_7 + b_7$ | 5318 MeV |
| t | $5_0 + 7_0 + 7_5 + 8_0 + 8_2$ | $u_5 + q_7 + t_7 + q_8 + t_8$ | 176.5 GeV |

The calculated masses are in good agreement with the observed constituent masses of ordinary leptons and quarks [9,10]. The mass of the top quark found by Collider Detector Facility is 176 ± 13 GeV [9] in a good agreement with the calculated value, 176.5 GeV. The masses of ordinary elementary particles can be calculated with only four known constants: the number of the extra spatial dimensions in the supermembrane, the mass of electron, the mass of Z°, and $\alpha_e$. The calculated masses are in good agreement with the observed values.

## 6. The Cosmic Connection

The cosmic vacuum and the four dimensional ordinary vacuum are connected by the cosmic connection everywhere all the time. When the spacetime dimensions in the two vacuums are different, the cosmic connection remains dormant, and the vacuum energy (cosmological constant) with respect to the observable universe is zero. When the spacetime dimensions in the two vacuums are compatible, the cosmic connection becomes active. When the cosmic vacuum is in the expansion phase, the expansion of the cosmic vacuum aids the expansion of the observable universe. The vacuum energy is positive. The positive vacuum energy causes negative pressure, inducing anti-gravity input and accelerating expansion. When the cosmic vacuum is in the contraction phase, the vacuum energy is negative. The negative vacuum energy causes positive pressure, inducing gravity input and accelerating contraction.

During one cycle of the universe, there are four periods of active cosmic connection: the inflation before the big bang, the late-time cosmic accelerating expansion, the cosmic accelerating contraction, and the final cosmic accelerating contraction. The first period of active cosmic connection is the inflation induced by anti-gravity input at the start of the expansion of the universe. The period is between the eleven dimensional Planck supermembrane and the eleven dimensional fermion. From Table 1 and Eq. 2, the energies of the Planck supermembrane and the eleven dimensional fermion are $1.1 \times 10^{19}$ GeV and $8.3 \times 10^{16}$ GeV, respectively. The energy ratio between the Planck supermembrane and the eleven dimensional fermion is $1.4 \times 10^{2}$. In other words, there are $1.4 \times 10^{2}$ eleven dimensional fermions per Planck supermembrane. The fractionalization rate is one particle per one-fourth power of the Planck time ($1.52 \times 10^{-11}$ second) per parent Planck supermembrane. The total time for the fractionalization is $2.1 \times 10^{-9}$ second that is the time for the inflation before the big bang (the cosmic instant fractionalization). As soon as the cosmic instant fractionalization takes place, the cosmic connection becomes dormant due to the difference in the spacetime dimensions between the two vacuums.

The dormant cosmic connection becomes active when the energy level in the cosmic vacuum is equal to the energy level between the five-dimensional fermion and the four dimensional boson. The vacuum energy in the observable universe again starts to become positive, inducing the negative pressure and anti-gravity input, causing accelerating expansion. It is the second period of active cosmic connection, which is happening at the present as late-time cosmic accelerating expansion.

The evidence for late-time cosmic accelerating expansion is from the recent observations of large-scale structure that suggests that the universe is undergoing cosmic accelerating expansion, and it is assumed that the universe is dominated by a dark energy with negative pressure recently [11]. The dark energy can be provided by a non-vanishing cosmological constant or quintessence [12], a scalar field with negative pressure. However, a cosmological constant requires extremely fine-tuned [13]. Quintessence requires an explanation for the late- time cosmic accelerating expansion [14]. Why does quintessence dominate the universe only recently?  One of the explanations is the *k*-essence model where the pressure of quintessence switched to a negative value at the onset of matter-domination in the universe [14].

According to the cosmic connection model, this late-time cosmic accelerating expansion is caused by the anti-gravity input during the second period of active cosmic connection.  The compatible energy level for the cosmic vacuum and the ordinary vacuum is between the energy levels of the five dimensional fermion and the four dimensional boson.  From Eqs (1) and (2) and Table 1, the energies of the five dimensional fermion and the four dimensional boson are calculated to be $2.72 \times 10^{-8}$ GeV and $1.99 \times 10^{-10}$ GeV, respectively.  The energy of the Planck supermembrane is $1.1 \times 10^{19}$ GeV from Table 1.  The energy ratio between the Planck supermembrane and the five dimensional fermion and the four dimensional boson are $4.17 \times 10^{26}$ and $5.71 \times 10^{28}$, respectively.  In other words, there are $4.17 \times 10^{26}$ five dimensional fermions and $5.71 \times 10^{28}$ four dimensional bosons per Planck supermembrane.  The fractionalization rate is one particle per one-fourth power of the Planck time ($1.52 \times 10^{-11}$ second) per parent Planck supermembrane.   The total time for the fractionalization is 0.2 billion years for the five dimensional fermion, and 27.6 billion years for the four dimensional boson.  Therefore, anti-gravity input starts in 0.2 billion years after the Big Bang, and ends in 27.6 billions years, and anti-gravity input causes accelerating expansion in the observable universe.

If the relative rate of the rise and the fall of anti-gravity input in the observable universe during this period follows a smooth curve with respect to time, the maximum relative rate (100) of anti-gravity input occurs in about 13.9 billion years after the big bang.  It is coincidentally about the age of the universe.  The relative rate of anti-gravity input versus time is shown in Fig. 2.

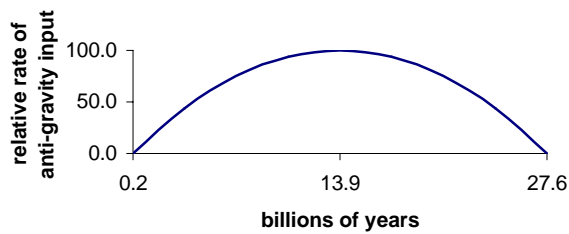

**Fig. 2.**  relative rate of anti-gravity input in the big bang universe versus time

In other words, the greatest acceleration occurs at the peak of the space absorption in 13.9 billion years after the big bang.

The oscillation of the cosmic vacuum now is in the expansion mode.  The cosmic vacuum expand only down to the beginning of the four dimensional spacetime, that is about 14 billion years from now.  When the oscillation is in the contraction mode in the future, and the cosmic vacuum and the ordinary vacuum are compatible in spacetime dimension, the observable universe will gain gravity input.  The result is the accelerating contraction of the observable universe.  This third period of the active cosmic connection prevents the observable universe to expand forever with an infinite large space.  Therefore, the observable universe and the cosmic vacuum can contract together.

Toward the end of the contraction mode, black holes become the dominating feature in the observable universe.  Black hole contains eleven dimensional black hole gluon and gluino surrounded by gravity.  At this point, even the ordinary vacuum becomes eleven dimensional.   When both cosmic vacuum and the ordinary vacuum are near the eleven dimensional spacetime, they become compatible.  In this last period of active cosmic connection, the observable universe again gains gravity input.  The result is to gather all material in the observable universe for the final contraction into the eleven dimensional black hole gluons and gluinos surrounded by four dimensional gravity in one area.

Moment after the final accelerating contraction, the cosmic vacuum starts the expansion phase, and anti-gravity input starts to start the inflation.  The anti-gravity input annihilates gravity surrounding the black hole gluons and gluinos.  Without gravity, black hole gluons and gluinos disappear as quarks disappear outside of hadrons.  What remain are the eleven dimensional Planck supermembranes that hold together by the tangle of D-branes and strings in the inflated space.  Immediately, the eleven dimensional Planck supermembranes undergo cosmic instant fractionalization, and another cosmic cycle starts.  The inflation, therefore, is not an add-on but a necessity to start the cosmic cycle.  Through different routes, both the cosmic vacuum and the observable universe achieve the cyclic fractionalization-condensation.

### *6.    Conclusion*

In this paper, quantum mechanics is explained by its cosmic origin. The reason for the existence of quantum mechanics is the spontaneous tendency for a microscopic object to fractionalize instantly into quasistates and condense instantly quasistates. This quasistate is equivalent to the eigenfunction in the wavefunction.  An object with the fractionalization-condensation is equivalent to the unitary wavefunction. Nonlocal operation is explicitly required to maintain communication among all quasistates regardless of distance during the fractionalization process.  Interference effect is explicitly required for the condensation of quasistates.  The collapse of the fractionalization-condensation is explicitly required when the fractionalization-condensation is disrupted externally by the interaction with environment and measurement or internally by transformation such as decay.  The object without fractionalization-condensation is the classical object without nonlocality-interference.

The cosmic origin of quantum mechanics is derived from the cyclic fractionalization-condensation in the cyclic universe, consisting of the unobservable cosmic vacuum and the observable universe.  The cyclic fractionalization-condensation allows quasistates to appear cyclically rather than simultaneously.  The unobservable cosmic vacuum involves the gradual cyclic fractionalization-condensation between the high energy eleven dimensional spacetime and low energy four dimensional spacetime.  The observable universe involves the drastic cyclic fractionalization-condensation consisting of the cosmic instant fractionalization (the big bang) into various dimensional particles and the expansion-contraction by mostly cosmic radiation and gravity.  The big bang (cosmic instant fractionalization) in the observable universe leads to the microscopic instant fractionalization-condensation that allows all quasistates from an object to appear simultaneously.  The microscopic instant fractionalization-condensation is the standard quantum mechanics.

The cosmic origin of symmetry is supersymmetry (the symmetry of spacetime) in the cyclic universe.  The supersymmetry leads to the eleven dimensional Planck supermembrane. The big bang generates missing space dimensions that are filled by internal space dimensions with internal symmetries.

The products of the big bang are the four spacetime dimensional ordinary (baryonic) matters and fields with seven dimensional internal space (non-spacetime), the exotic dark matters with higher spacetime dimensions, and the four spacetime dimensional gravity, cosmic radiation, and ordinary vacuum without internal space. The masses of ordinary elementary particles can be calculated with only four known constants: the number of the extra spatial dimensions in the supermembrane, the mass of electron, the mass of $Z°$, and $\alpha_e$. The calculated masses are in good agreement with the observed values.

The cosmic connection between the cosmic vacuum and the ordinary vacuum becomes active, when the spacetime dimensions of the cosmic vacuum and the ordinary vacuum are compatible.  The active cosmic connection allows the observable universe to have expansion input or contraction input from the expansion or contraction of the cosmic vacuum.   The results are the inflation before the big bang, the late-time cosmic accelerating expansion, the cosmic accelerating contraction, and the final cosmic accelerating contraction. Through different routes, both the unobservable cosmic vacuum and the observable universe achieve the cyclic fractionalization-condensation.